# A microfluidic platform for actin-based membrane remodeling reveals the stabilizing role of branched actin networks on lipid microdomains.

Lixin Huang[1], Rogério Lopes Dos Santos[1], Sid Labdi[1], Guillaume Lamour[1], Olek Maciejak[1], Michel Malo[1], John Manzi[2], Martin Lenz[3], Jacques Fattaccioli[4,5] Clément Campillo[1,6]*

[1] Université Paris-Saclay, Univ Evry, CY Cergy Paris Université, CNRS, LAMBE, 91025 Evry-Courcouronnes, France;

[2] Institut Curie, Université PSL, Sorbonne Université, CNRS UMR168, Physique des Cellules et Cancer, 75005 Paris, FR.

[3] LPTMS, CNRS, Université Paris-Sud, Université Paris-Saclay, 91405 Orsay, France.

[4] CPCV, Département de Chimie, École Normale Supérieure, PSL University, Sorbonne Université, CNRS, 75005 Paris, France;

[5] Institut Pierre-Gilles de Gennes pour la Microfluidique, 75005 Paris, France.

[6] Institut Universitaire de France (IUF), 75231 Paris, France;

* Correspondence to: clement.campillo@univ-evry.fr

Authors email:

L.H.: lixin.huang@etu-upsaclay.fr

R. LDS.: rogerio.lopesdossantos@univ-evry.fr

S. L.: sidahmed.labdi@univ-evry.fr

G. L.: guillaume.lamour@univ-evry.fr

O. M.: olek.maciejak@univ-evry.fr

M. M.: michel.malo@univ-evry.fr

J.M.: John.Manzi@curie.fr

M.L.: martin.lenz@cnrs.fr

J. F.: jacques.fattaccioli@gmail.com

C.C.: clement.campillo@univ-evry.fr

**Abstract**

Cell shape changes, essential for processes such as motility or division, are controlled by the actomyosin cortex that actively remodels biological membranes. Their mechanisms can be deciphered *in vitro* using biomimetic reconstituted systems, such as giant unilamellar vesicles (GUVs) with controlled lipid composition coupled to reconstituted actin networks. These assays allow mimicking cell shape changes in controlled biochemical and biophysical environments. However, studying the dynamics of these shape changes on statistically significant populations of GUVs with the possibility to sequentially modify the protein composition of the assay is a major experimental challenge. To address these issues, we use a microfluidic approach to immobilize several dozens of isolated GUVs and monitor membrane and actin network evolution. We first characterize the loading of the chamber with GUVs and actin. Then, we monitor the actin-induced remodeling of populations of homogeneous and phase-separated GUVs and show that actin networks prevent the coalescence of lipid microdomains and that, in return, the number of domains affects the actin network structure. Our microfluidic-based experimental strategy thus allows for studying actin-induced membrane deformation *in vitro* and could be adapted to other studies on membrane remodeling processes.

# 1 Introduction

The bottom-up approach in synthetic biology uses *in vitro* model systems to investigate complex biological processes, such as cell shape changes that occur in cell migration.[1,2] This approach relies on the use of controlled assays with known biochemical environments. Giant Unilamellar Vesicles (GUVs) are biomimetic cell-sized membrane spheres encapsulating aqueous solutions. GUVs are robust models mimicking biological membranes' lipid and protein composition.[3] Hence, they are well suited to the *in vitro* reconstitution of biological processes such as endocytosis,[4,5] motility,[6,7] or intercellular communication.[8]

Cell shape changes result from the interplay between membranes and cytoskeleton dynamics.[9] The actin cortex, which underlies the plasma membrane, is a major player in these cell shape changes by generating forces on the membrane through actin polymerization and myosin activity.[10,11] Actin dynamics are regulated by many Actin-Binding Proteins. Profilin accelerates actin-bound nucleotide exchange and promotes actin polymerization.[12,13] Capping protein stabilizes actin filaments by impeding assembly and disassembly at the filament barbed end.[14] Actin polymerization is activated by Nucleation Promoting Factors (NPFs) as proteins of the WASp (Wiscott-Aldrich Syndrom protein) family. They are located at the plasma membrane and comprise a VCA (Verprolin homology-central-acidic region) domain that activates the Arp2/3 complex[15], initiating the nucleation of new filaments from the side of pre-existing filaments and thus forming branched F-actin networks.[16]

*In vitro*, dynamic actin networks based on the above proteins can be reconstituted in the vicinity of lipid membranes. These experimental assays have allowed the reproduction of various cell-mimicking structures made of actin and membranes[17], including filopodia-like protrusions[18–20], nanotubes interacting with actin networks[21–23], blebs,[24] or cytokinetic rings.[25] Branched Arp2/3 actin network polymerization can be triggered at the outer surface of GUVs by attaching NPF to their membrane[18,26,27]. In these experiments, as in the case of rigid beads coated with NPF[28,29], the actin network grows mainly by the addition of actin monomers at the surface of the GUV. At the onset of the actin polymerization, this leads to a symmetric shell of branched Arp2/3-nucleated actin gel surrounding the GUV (or bead). As monomers continue being added close to the GUV surface, this actin shell induces compressive forces toward the GUV center and is submitted to increasing mechanical stress. After a certain threshold, these polymerization-induced forces lead to a localized rupture of the actin gel, inducing a symmetry-breaking of the whole system.[30] From this step, the actin shell is asymmetric. Finally, NPFs get polarized on one side of the GUV, and actin monomers continue to be added on this side, forming an actin "comet" that propels the GUVs by a mechanism reminiscent of the propulsion of *Listeria Monocytogenes*.[31]

Three experimental challenges remain to be addressed in studies on GUVs coupled to actin networks: sequentially adding proteins onto immobilized GUVs, following single GUVs over time to study their actin-induced remodeling, and obtaining statistics over GUV populations. Overcoming these bolts would allow sophisticated *in vitro* experiments to decipher the complex mechanisms that control cell membrane remodeling by actin and other proteins.

Besides, biological membranes are highly heterogeneous and include the formation of lipid domains, which affect various cellular processes such as immune response.[32,33] In liquid-liquid phase-separated membranes, sphingolipids and sterols having longer and more saturated hydrocarbon chains form liquid-ordered (Lo) phases, while unsaturated lipids form liquid-disordered (Ld) phases.[34] How the actin cytoskeleton interacts with these liquid domains is unclear. Many *in vivo* experiments have shown a crosstalk between actin dynamics and membrane organization. Examples include the assembly of WASH (an NPF belonging to the WASp family) domains on endosomes,[18] the formation of membrane nanodomains on membrane tubules,[35] or the formation of Lo nanodomains induced by actin attachment[36], thus suggesting that actin dynamics could favor the apparition of lipid domains. Besides, *in vitro* studies have shown that actin networks influence the demixing temperature, location,[37] and number of domains.[38] Nevertheless, the role of actin dynamics in the formation, maintenance, or coalescence of membrane domains remains unclear. From the synthetic cell perspective, lipid domains are a proxy to control the spatial localization of biochemical processes at the surface of lipid membranes.[38,39] Therefore, novel tools are required to investigate the spatiotemporal crosstalk between actin dynamics and lipid domains.

Microfluidic systems enable the production,[40] sorting,[41] and confinement of GUVs.[42] They have also been employed to visualize the separation and coalescence of GUV lipid domains.[43] Nevertheless, a microfluidic approach has yet to be developed to study actin-induced remodeling. In this study, we develop a strategy to follow GUV remodeling driven by branched Arp2/3-nucleated F-actin networks in PDMS microfluidic chambers. We have developed a passivation protocol that prevents both protein and lipid adsorption on the PDMS. Our platform allows a large population (typically a hundred GUVs) to be loaded and immobilized in the microfluidic chamber in a few minutes. A quarter of them are isolated in a trap, thus suitable for membrane remodeling experiments. This platform first enabled us to reproduce existing results on homogeneous GUVs, validating our experimental strategy. Then, we used it to study the coupling between the actin network and phase-separated GUVs. We demonstrate not only the effect of the number of membrane domains on the actin network structure but also the actin-induced stabilization of membrane domains, providing new insights into the effect of actin on membrane dynamics and vice-versa.

## 2 Methods

**Lipids, proteins, and reagents**

1,2-Dioleoyl-sn-glycero-3-phosphocholine (referred to as DOPC), N-(dodecanoyl)-sphing-4-enine-1-phosphocholine (sphingomyelin, referred to as SM), cholesterol (referred to as Chol) from ovine wool, DSPE-PEG (2000)-Biotin (1,2distearoyl-sn-glycero-3-phosphoethanolamine-N-[biotinyl-(polyethylene glycol)2000]) (PEG-biotin lipids), and 18:1 DGS-NTA(Ni)(1,2-dioleoyl-sn-glycero-3-[(N-(5-amino-1-carboxypentyl) iminodiacetic acid)succinyl] (Ni-NTA) are obtained from Avanti Polar Lipids. Fluorescent Texas Red r-1,2-dihexadecanoylsnglycero-3-phosphoethanolamine triethylammonium salt (Texas-Red DHPE) lipid is purchased from Thermo-Fisher. Polydimethylsiloxane (PDMS) and curing agent are obtained as SYLGARD 184 silicone elastomer kit from Darwin Microfluidics. Poly(L-lysine)-graft-poly(ethylene glycol) (PLL-PEG) pellets are provided by SuSoS. Pluronic F-127 pellets, Polyvinylpyrrolidone (PVP) pellets, Tris, β-casein, ATP, KCl, $MgCl_2$, $CaCl_2$, dithiothreitol (DTT), NaOH, 99.8% ethanol and sucrose are obtained from Merck. Actin from rabbit muscle, porcine Arp2/3 complex, and recombinant Human Profilin-1 are purchased from Cytoskeleton. Label-free Capping protein (CP) and Actin-Atto 488 conjugate are from Hypermol. Streptavidin-pVCA-Histidine (SpVCA-His, where pVCA is the proline-rich domain-verprolin homology-central-acidic sequence from human WASP, starting at amino acid Gln150) that we call spVCA in this paper is a gift from UMR 168, PhysicoChimie Curie.[44]

**Buffers**

GUVs electroformation buffer contains 200 mM sucrose and 2 mM Tris, adjusted to pH 7.4 and 200 mOsm (Löser Micro-Digital Osmometer Type 16i, Camlab). The polymerization buffer contains 165 mM sucrose, 1 mM Tris, 50 mM KCl, 2 mM $MgCl_2$, 0.1 mM DTT, 10 mg/mL β-casein, and 2 mM ATP in Milli-Q water and delicately adjusted at 7.4. For high protein concentration assays, the polymerization buffer is set and measured at 278 mOsm; when protein solution is added, the solution's osmolarity is finally 200 mOsm. We prepare a solution containing 30 μM G-actin with 15% labeled Atto 488 actin conjugate. Then, this solution is incubated in G-Buffer (2 mM Tris, 0.2 mM $CaCl_2$, 0.2 mM ATP, 0.2 mM DTT, pH 8.0) overnight at 4 °C to ensure F-actin depolymerization.

**GUV preparation**

Our technique for GUV fabrication is electroformation.[45] For electroformation, the lipid concentration is 2.5 g/L in the solvent chloroform/methanol 5:3 (v/v). The lipid composition of DOPC GUVs incorporates 94.5% DOPC, 5% Ni-NTA lipid (or 1% PEG-biotin lipids with 98.5% DOPC), and 0.5% Texas-Red DHPE lipids. The tertiary phase-separated lipid composition of GUVs includes DOPC/SM/Chol at a molar ratio of 2/2/1, plus 2% Ni-NTA and 0.5% Texas-Red DHPE lipids. Using cleaned 10 μl Hamilton syringes, a 5 μl lipid mixture was spread on each conductive surface of ITO glass slides (Indium Tim Oxide glass slides, Merck). A framed electroformation chamber was formed using the ITO slides, with conductive sides facing each other. They are sealed with a hollow rectangular cured PDMS frame, and the 2 ITO slides are clamped. The framed electroformation chamber is then injected with 0.4 mL of electroformation buffer. GUVs are formed by applying an alternate sinusoidal current (10 Hz, 3 V peak to peak) for 2 hours. DOPC GUVs are electroformed at room temperature. Phase-separated GUVs are placed inside a convection oven prewarmed at 60 °C, beyond their demixing temperature.[46,47] The resulting GUVs are carefully collected in a 1.5 mL Eppendorf tube filled with filtered nitrogen. GUVs solution is stored at 4 °C in the dark to prevent photobleaching of Texas-Red and can last for one week.

**Microfluidic devices fabrication**

Microfluidic devices were made of PDMS (polydimethylsiloxane), using standard soft lithography techniques.[48] We fabricated SU-8 master molds (Microchem) on a single-side polished silicon wafer after designing microtrap arrays in a microfluidic chamber. The patterned silicon wafer is then replicated by epoxy, for repeat use. Microfluidic chips are prepared by mixing PDMS elastomer with a curing agent at a ratio of 10:1 (w/w) at room temperature. The mixture is sonicated for 5 minutes and degassed to remove air bubbles (Millivac Mini vacuum pump, Merck Millipore). The degassed mixture is then poured over the patterned epoxy replicate and cured at 80°C overnight. Cured PDMS is peeled off the replicate mold and cut into individual chips, and fluidic inlet/outlet are punched out using commercial biopsy punchers (Kai Medical Biopsy Punches, Meidca100, 2.5 mm diameter for inlet channel, and 1 mm for outlet channel). The inlet channel is a small reservoir with a capacity of about 65 μl. The microfluidic chip and coverslips (Menzel Gläser Coverslips 24 mm×50 mm, Fisher Scientific) are sonicated in 1 M NaOH solution for 15 minutes and then placed in a clean hood for 24 hours.[49] PDMS chips and coverslips are rinsed with Milli-Q water and dried with filtered gas nitrogen in a clean hood. To assemble the microfluidic device, both the PDMS chips and the glass coverslips are activated with $O_2$ plasma (Plasmaflo PDC-FMG-2 and Plasma Cleaner PDC-32G-2, Harrick Plasma) at 150 mTorr (about 20 Pa) for 2 min. Right after the activation step, they are aligned by hand, put gently into contact, and heated for 1 min at 100 °C on a hot plate. A second $O_2$ plasma treatment on the microfluidic assembly is operated at 70 mTorr for 10 min. After, the microfluidic assembly is immediately incubated in 5% 10 kDa PVP (w/v in dd water) passivation solution and placed on an orbital shaker for one night. Before microfluidic experiments, the microfluidic assembly is retrieved and stirred in Milli-Q water for 1 hour.

**Nucleation-promoting factors (NPFs) incubation and actin polymerization**

SpVCA is diluted in the electroformation buffer at a final concentration of 788 nM. We prepare 3 μl of spVCA solution and aspire it for 5 minutes at an imposed flow rate of 0.5 μl/min. The flow rate is then maintained at 0.05 μl/min, providing a quasi-static incubation of spVCA. Finally, actin and ABPs solution that comprises 6 μM actin, 6 μM profilin, 50 nM capping protein, and 80 nM Arp2/3 complex in the polymerization buffer is pipetted into the inlet reservoir. We run the protein introduction for 5 minutes at a 0.5 μl/min flow rate. We reduced it to 0.05 μl/min for quasi-static actin polymerization.

To study phase-separated GUVs, the protocol is adapted as follows. After electroformation at 60°C, the ternary lipid components in the GUVs are fully miscible, leading to a homogeneous distribution of the fluorescent lipids. When the temperature drops below 37°C, immiscible domains emerge as small nuclei, rapidly coalescing into larger domains. Texas-Red DHPE and Ni-NTA lipids enrich the unsaturated Ld domains after the lipid demixing transition.[38] Since spVCA activates Arp2/3 complex and triggers actin polymerization, the resulting polymerizing actin network colocalizes with the Ld fluorescent lipid domain, and can be monitored over time.

**Bright-field and fluorescence imaging**

Lipids and proteins are imaged using a Zeiss Axio observer 7 inverted epifluorescence microscope (GFP filter cube, excitation 470 nm, emission 525 nm; Texas

Red filter cube: excitation 545−580 nm, emission 615 nm) equipped with either a 10X air or a 63X-1.4 numerical aperture oil-immersion objective. Images are collected by an AxioCam MRm camera. For all actin polymerization experiments, the exposure times are 22 and 30 milliseconds for DHPE-Texas Red (lipid membrane) and Actin-Atto 488 conjugate (actin network). The fluorescent probes are excited using a mercury plasma light, and fluorescence emissions were collected for Texas Red DHPE and Actin Atto 488 with excitation/emission wavelength band-passes of 596 nm/615 nm and 495 nm/518 nm, respectively. Image acquisition is controlled by Zeiss Blue software from Zeiss Group, and image analysis was conducted using Fiji. For bulk actin fluorescence quantification, we used the Zeiss *Apotome* module, enabling a confocal-like single-layer fluorescent signal acquisition.

We perform several fluorescence quantifications on the acquired images. First, GUVs are delineated using the Fiji *Wand Tool*. The GUV diameter is defined as:

$$d_{GUV} = 2\sqrt{\frac{Area_{GUV}}{\pi}}$$

Where $Area_{GUV}$ is the value provided by Wand Tool.

The Ellipticity of GUVs or actin networks is defined as:

$$Ellipticity(\%) = \frac{L_{Max\_Perp.}}{Maximum\ Feret\ Diameter} \times 100$$

Where *Maximum Feret Diameter* is the maximum distance of the 2D contour of either GUV membrane or branched actin network, and $L_{Max\_Perp.}$ is the maximum contour distance that is perpendicular to the direction of *Maximum Feret Diameter*.

The actin thickness is measured by the Fiji *Length* tool as the longest radial distance between the GUV membrane and the edge of the actin network. Symmetry-breaking orientations are manually measured using the Fiji *Angle* tool. To measure the radial fluorescence along the GUV perimeter *(Fig-5B)*, we first manually determine the center of each GUV. Then, we draw radial profiles every degree by rotating around the center. We subtract the background intensity on each fluorescence channel. Then, for each radial profile, we measure the mean fluorescence intensity value on the peak corresponding to either the lipid membrane or the actin network. Lipid or actin fluorescence along the GUV perimeter is normalized: 1 corresponds to each channel's maximum and 0 to the background. Finally, we quantify the actin fluorescence on Ld domains in phase-separated GUVs. To do so, we arbitrarily define Ld domains as regions where the lipid signal is superior to 0.2 according to the quantification above. We measure the actin fluorescence for all the points in these regions and calculate the average actin fluorescence value on Ld domains per GUV. These average values are normalized between 1, corresponding to the maximal actin fluorescence on the Ld domains on a single GUV, and 0, the background.

Image analysis is carried out by MATLAB Simulink, and the statistical analysis is done with RStudio. For statistical significance test, we apply the following standards: n.s. $p > 0.05$; *$p \leq 0.05$; **$p \leq 0.01$; ***$p \leq 0.001$. All p-values are given in *Supplementary Tables S2&S3*.

## 3 Results and Discussion

**Experimental strategy**

We develop a microfluidic approach to monitor actin polymerization on GUVs using fluorescence microscopy. Figure 1A shows the design of the microfluidic chamber, which is 100 μm deep and has a volume of approximately 2 μl. It contains 10 rows of C-shaped microtraps (112 traps in total), each composed of four micropillars (Fig. 1B).[50,51] These microtraps facilitate GUV entrapment by guiding fluid streamlines between the micropillars. The inlet and outlet channels are offset to ensure homogeneous GUV loading. A syringe pump connected to the outlet generates aspiration pressure, enabling controlled filling of the MFC. PDMS is hydrophobic[49] and may adsorb proteins and GUV membranes. To avoid this, we first needed to passivate the PDMS. We tried Poly-L-lysine-graft-polyethylene-glycol (PLL-PEG, 22kDa) and Pluronic F-127 without success (*Fig-S1*). Finally, we found that polyvinylpyrrolidone (PVP) at 5% (m/V) effectively prevents protein and membrane absorption. This follows the published values of surface hydrophobicity of PDMS (Supplementary Table S1).

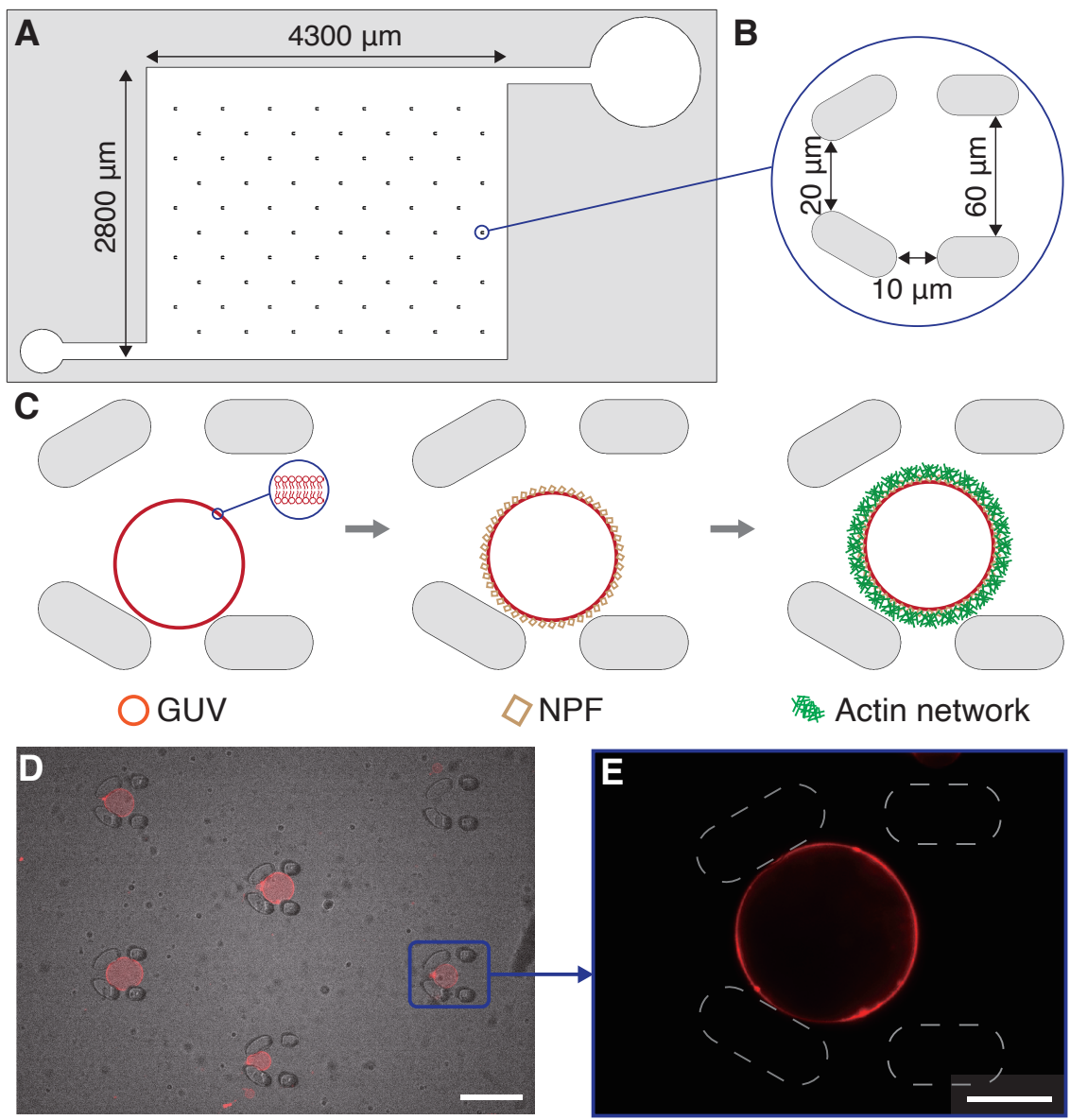

**Figure 1: Design of the microfluidic device and experimental strategy.** (A) Scheme of the microfluidic chamber, where the larger circle on the right is the inlet channel, and the smaller one on the left the outlet; (B) *C-like* traps composed of four micropillars; (C) 3-step experimental strategy: GUV filling, incubation of NPFs on the GUVs, and branched Arp2/3-nucleated F-actin network polymerization. (D) GUVs loading and entrapment: Bright-field and epifluorescence images of DOPC GUVs in microtraps, scale bar: 100 µm;

(E) Close-up of an individual trapped GUV, scale bar: 20 μm;

(*Fig-1C*) presents our experimental strategy: we first introduce GUVs into the microfluidic chamber. Then, we inject the NPF "spVCA", which comprises two binding tags, streptavidin and histidine, which allow its binding to either biotinylated or nickel (Ni-NTA) lipids. Finally, we add the actin and actin-binding proteins solution, and actin polymerization starts at the GUV membrane. In the following, we present the details of these experimental steps.

First, we study the loading of GUVs in the microfluidic chamber. We record bright-field and fluorescence images of the whole chamber (*Fig-1D&E*) every 3 minutes during GUVs filling and quantify the fraction of traps occupied by GUVs (*Supplementary Fig-S2A&B*). GUVs are introduced at a flow rate of 0.5 μl/min, both fast enough to ensure a fast filling of the chamber and slow enough to prevent GUVs from slipping out of the microtraps. We quantify the occupancy of the microtraps: 100% occupancy corresponds to all 112 traps filled with at least one GUV. This fraction is reached in approximately 30 minutes (*Supplementary Fig-S2A*). For actin experiments, we focus on isolated GUVs. This is the case for 25% of the microtraps (*Supplementary Fig-S2B)*. Therefore, in 30 minutes, the MFC is filled with dozens of individually trapped GUVs ready for further experiments. We then quantify the actin concentration in the microfluidic chamber by measuring the fluorescence intensity of solutions of fluorescently labeled G-actin. (Supplementary *Fig-S3*). Therefore, under our conditions, the chamber can be entirely and homogeneously filled with the actin cocktail in 5 min.

**Actin polymerization on GUVs in microfluidic chambers**

After these characterizations, we compare F-actin polymerization on GUVs in microfluidic chambers and conventional coverslips-made chambers. As already shown,[29,38,52,53] after 15 to 20 minutes of actin polymerization in closed coverslips, three sorts of actin network structures are observed on GUVs: symmetric actin shells, asymmetric actin shells, and actin comets (*Fig-2A*). These correspond to successive steps in the process of branched Arp2/3 actin network growth around spherical objects such as GUVs or rigid beads coated with NPFs.[26,54]

We compare the distribution of these three actin architectures as a function of the GUV diameter in microfluidic chambers and conventional closed coverslips (*Fig- 2B&C, SuppFig S2*). In both cases, symmetry breaking of the actin shells occurs when actin network thickness reaches 5-10 μm, and the network structure depends on the GUV size: GUVs under 20 μm are more likely to form actin comets, and larger ones are more often observed at the asymmetric shell stage. Importantly, we obtain the same repartition of actin structures in microfluidic chambers and conventional chambers (*Fig-2B&C*). The only difference lies in the size distribution of the GUVs: all sizes are observed in coverslips, whereas the microtraps select GUVs between 10 and 60 μm (*Fig-2B&C*).

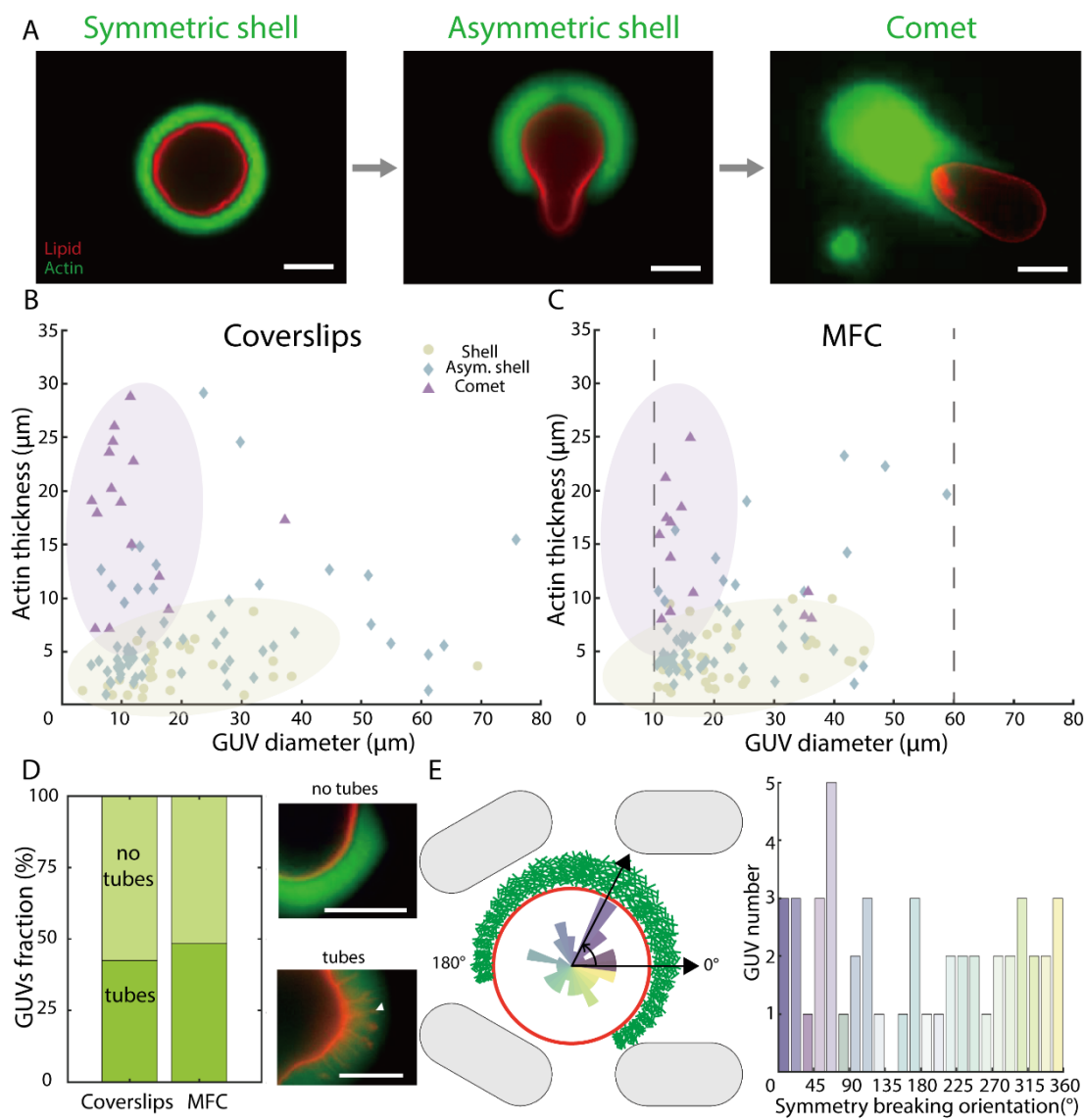

**Figure 2**: **Branched Arp2/3-nucleated F-actin networks polymerization on GUVs.** (A) Types of actin structures observed when polymerizing actin gels around GUVs: first, "actin shells" are observed, then symmetry breaking happens, forming "asymmetric shells", and later actin "comets" form; (B&C) Actin thickness as a function of GUV diameter in coverslips and MFC, showing the distribution of the three actin architectures, dashed lines highlight the GUV sizes filtered by microtraps (102 GUVs in MFC; 104 GUVs in closed coverslips; respectively from 14 and 16 independent experiments), violet and yellow areas are guides for the eyes, identical in B&C, and show regions where the majority of comets and symmetric shells respectively are observed ; (D) Fraction of polymerized GUVs containing membrane nanotubes in the actin network for coverslips and MFC (same population as in *(C)*); (E) Symmetry breaking orientation for GUVs in MFC (50 GUVs from N = 5 independent experiments). Scale bars: 10 µm.

Similar experiments in coverslips-made chambers have shown that actin polymerization dynamics generate membrane nanotubes embedded in the actin network.[20] Therefore, to further assess polymerization efficiency, in particular in terms of actin-membrane interaction, we compare the fraction of GUVs presenting membrane nanotubes in coverslips and microfluidic chambers and obtain the same proportion (*Fig-2D*). Finally, we checked that all symmetry-breaking orientations could arise in the traps (*Fig-2E*), implying that they do not clearly impose a defined orientation of the symmetry-breaking process.

Altogether, our results show that actin polymerization and membrane remodeling efficiency are identical in our microfluidic chambers compared to conventional coverslip-made chambers. Therefore, our approach, which combines GUV entrapment and controlled sequential protein addition, provides a robust system for studying the interplay between actin networks and GUV membranes, as illustrated below.

**Remodeling of GUVs with homogeneous membrane composition**

Here, we show that our microfluidic chamber allows monitoring actin-induced GUV deformations over time. At the beginning of the experiment, the positions of bare GUVs are localized, and their initial morphology is captured. We then inject the actin cocktail and track the GUV remodeling over time (*Fig-3A*). The actin gel thickness increases quasi-linearly with time (*Fig-3B*) at speeds ranging from 0.1 to 0.3 μm/min for symmetric and asymmetric actin shells, consistent with previous experiments on GUVs.[26] Moreover, in earlier bead-motility experiments, for beads coated with ActA[55] and N-WASp[54], the actin network growth rate decreases as the bead size increases: 0.9 μm/min for 10 μm diameter beads and 1.7 μm/min for 4.6 μm diameter beads in the case of N-WASP.[54] The actin growth rate also exhibits a size-dependent behavior in our experiments (*Fig-3C*). Altogether, these experiments show that we can capture branched actin network growth dynamics around homogeneous GUVs in microfluidic chambers.

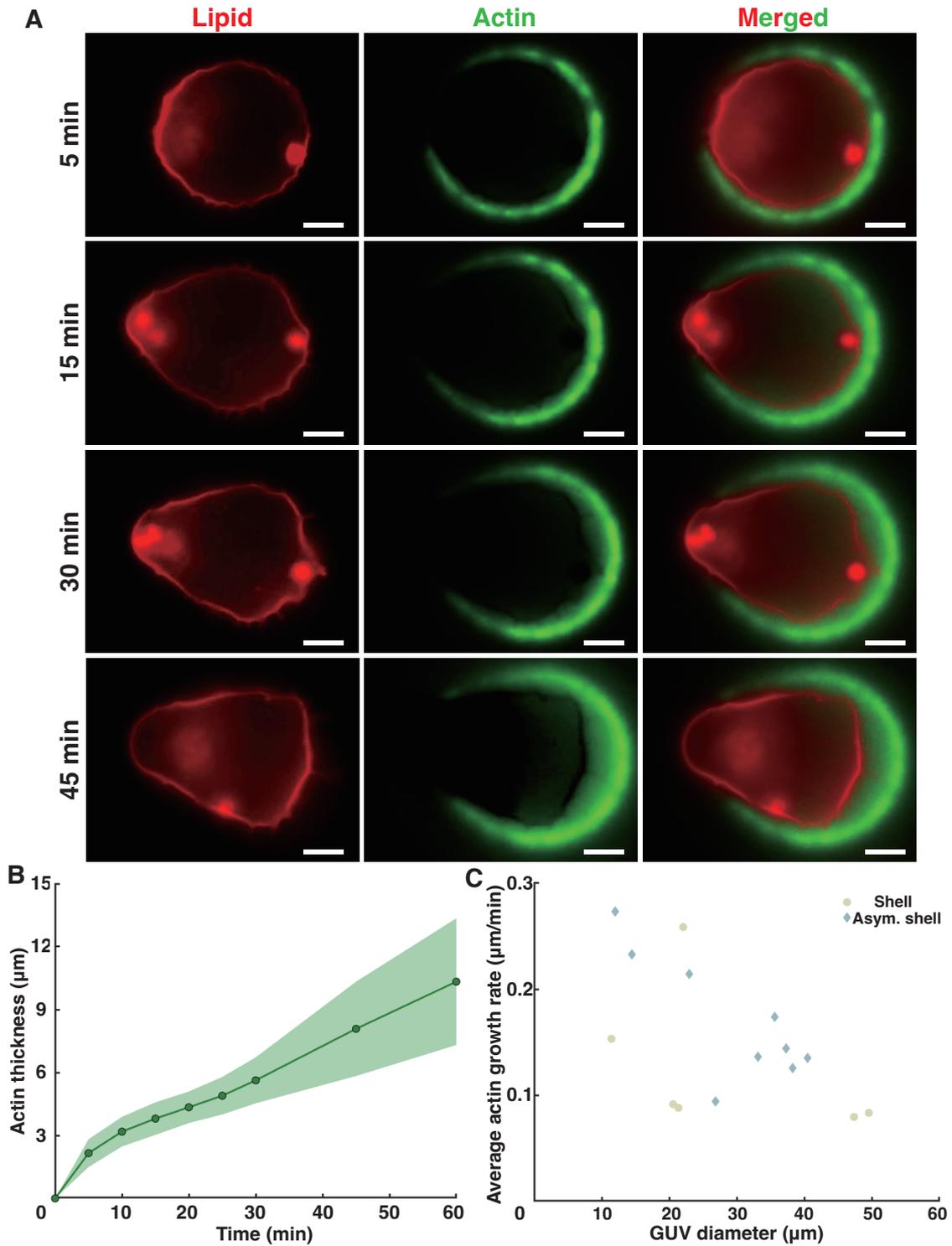

**Figure 3: Real-time tracking of homogeneous GUV remodeling.** (A) Time-lapse fluorescence microscopy images showing the progression of branched Arp2/3-nucleated F-actin networks polymerization over 1 hour; Scale bars: 10 μm (B) Actin thickness growth vs. time (mean ± standard deviation, N = 16 GUVs); (C) Mean actin network growing speed vs. GUV size (N = 16 GUVs).

**Actin networks stabilize lipid domains**

We now exploit our microfluidic chamber to assess how branched Arp2/3-nucleated F-actin networks polymerized on Ld domains affect lipid domain coalescence.

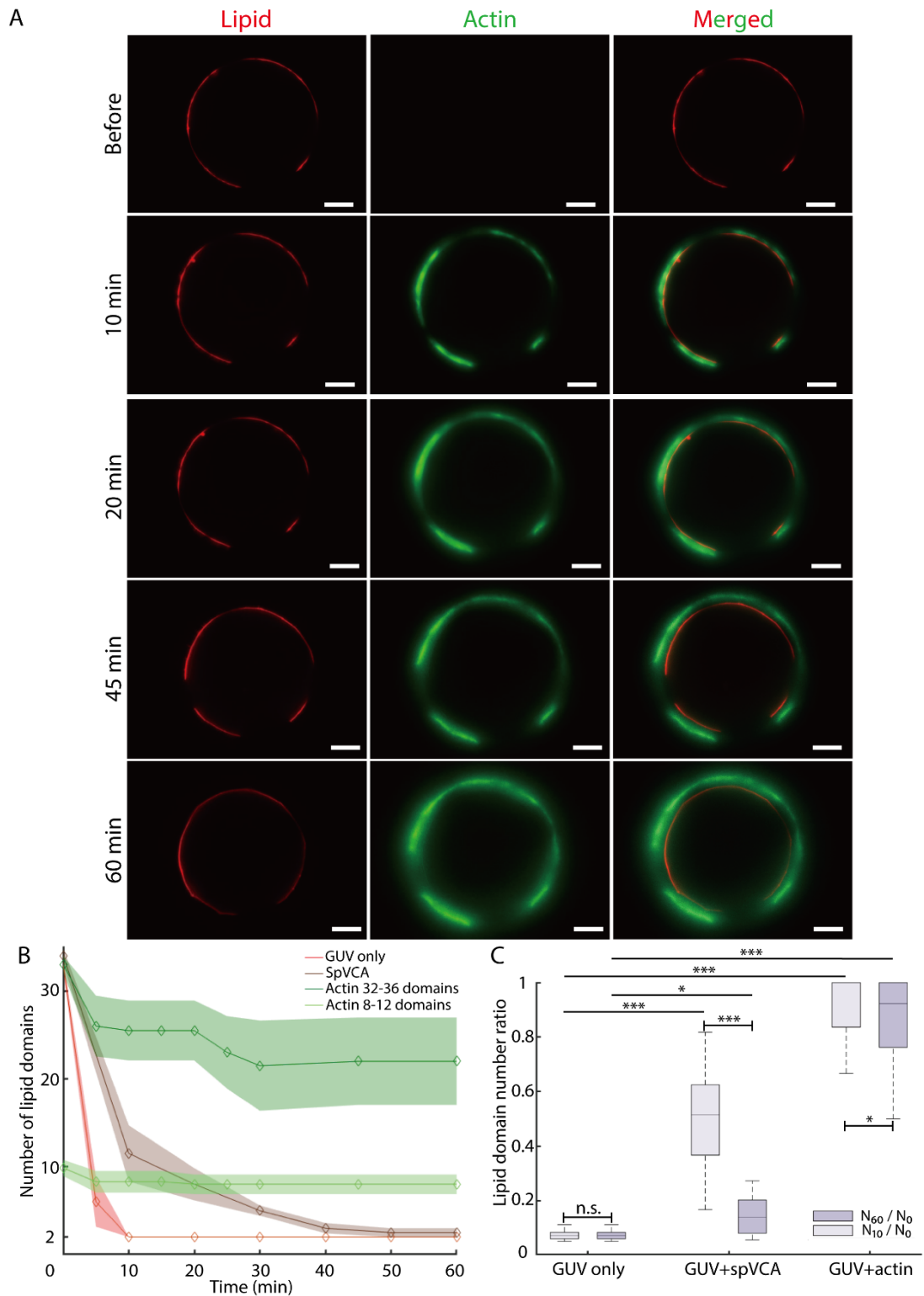

**Figure 4**: **Branched Arp2/3-nucleated F-actin networks stabilize lipid domains.** (A) Fluorescence microscopy time-lapse sequence showing the growth of an actin network on a phase-separated GUV Scale bar: 10 μm; (B) Coalescence dynamics of lipid domains on GUVs with initially 32-36 domains or 8-12 domains (mean ± standard deviation) in the equatorial focus plan (GUVs, N = 7; spVCA, N = 8; Actin 32-36 domains, N = 9; Actin 8-12 domains, N = 9); (C) Domain number ratio at 10 and 60 minutes after the lipid demixing transition, for bare GUVs, spVCA-bound GUVs, and GUVs with actin polymerization

(GUV, N = 19; spVCA, N = 20; Actin, N = 4). Statistical analysis in Fig 6C was performed using Mann−Whitney U tests among the three independent GUV configurations and Wilcoxon rank-sum tests between paired time parameters; non-parametrical tests are operated because samples differ from a normal distribution.

Here, we follow over time the growth of branched Arp2/3 nucleated actin networks on the Ld domains of phase-separated GUVs, adapting a developed protocol.[38] Multi-component GUVs are loaded into the MFC, maintained at 37.2°C. After 30 minutes of GUV loading, we introduce the spVCA solution at 37.2°C into the MFC. Following a 15-minute NPF incubation, we turned off the heating unit and finally introduced the actin and ABP solution, which was maintained at 4°C on ice. We indeed observe the growth of actin networks over time and their colocalization with fluorescent Ld domains (*Fig-6A*). We also conducted two control experiments: (a) Negative control without protein, where multi-component GUVs were pre-loaded into the MFC at 37.2°C for 30 minutes, followed by the injection of buffer at 4°C for 5 minutes, followed by 55 minutes of quasi-static flow. (b) Positive control with GUVs coated with spVCA, where the same GUV pre-loading process was followed by 15 minutes of spVCA incubation. Buffer was then injected in the same manner as in the negative control. Figure 6B shows the number of lipid domains over 1 hour for three conditions: bare phase-separated GUVs, GUVs coated with spVCA, and GUVs with branched Arp2/3-nucleated F-actin networks on Ld domains; for this latter case, for the sake of clarity, we present only two groups of GUVs according to their initial domain number, 32–36 or 8–12 domains. The results are similar for all other initial numbers of lipid domains. After electroformation at 60°C, the ternary lipid components in the GUVs are fully miscible, leading to a homogeneous distribution of the fluorescent lipids. When the temperature drops below 37°C, for bare GUVs, domains emerge as small nuclei, rapidly coalescing into larger domains. The number of lipid domains in bare GUVs drops to 2 within 10 minutes, indicating complete domain coalescence, consistent with previous results.[56] SpVCA reduces the coalescence speed in the positive control, with the domain number stabilizing at two or four after one hour. SpVCA binding attachment to the lipids might slow coalescence by reducing the Ld domains diffusion coefficient, an effect that should be even stronger in the case of small domains.[57,58] With actin polymerized on Ld domains, we observe a slight domain coalescence during the first 10 minutes of actin polymerization. Then, the domains are stabilized, regardless of the initial domain number. We have also checked that, after 3h, the domains are still stabilized. To quantify this stabilization, we define the "domain number ratio" as the number of domains at 10 or 60 minutes, $N_{10}$ or $N_{60}$, normalized by the initial domain number $N_0$ (*Fig-6C*). A high ratio indicates less lipid domain coalescence or a strong stabilization effect. Figure 6C shows that the ratio for bare GUVs is the lowest for 10 and 60 minutes, indicating complete domain coalescence. For spVCA-coated GUVs, the ratio is close to bare GUVs at 60 min (see detailed statistical test for *Fig-6C* in *Table-S1*). Yet, at 10 minutes, only half of spVCA- GUVs completed coalescence, confirming that the presence of SpVCA slows down the coalescence process. For GUVs with actin networks, as shown by the boxplots that highlight the quartiles of the considered populations, half of the population exhibited no domain coalescence within the first 10 minutes. After 1 hour, at least a quarter of them remained completely stabilized. These results show that branched Arp2/3-nucleated F-actin networks polymerized on Ld domains impede domain coalescence, leading to domain stabilization. These results could explain the *in vivo* experiments in which actin networks are associated with membrane microdomains: actin could stabilize these domains by preventing their coalescence.

**Lipid domain number influences actin network structure**

Here, we take advantage of the ability of our approach to generate GUVs with various domain numbers to assess how this parameter affects the structure of branched Arp2/3-nucleated F-actin networks polymerized on the Ld domains of GUVs. We define three categories of GUVs based on the initial number of lipid domains $n_d$: (a) $n_d = 2$; (b) $2 < n_d \leq 10$; (c) $n_d > 10$ *(Fig-5A)* and compare the actin networks at their surface based on actin fluorescence profiles along their perimeter *(Fig-5B)* and total actin fluorescence on Ld domains (*Fig-5C*).

For $n_d = 2$ and $2 < n_d \leq 10$, actin strongly colocalizes with Ld domains as in *Fig-6*, and the domains display the strongest actin fluorescence. For $n_d > 10$, the actin network appears almost homogeneous over Ld and Lo domains, and its fluorescence is lower compared to the two latter cases, demonstrating a lower actin density. Remarkably, even though the actin networks appear almost homogeneous, we never observed symmetry-breaking in the case of heterogeneous GUVs with $n_d > 10$ as observed for homogeneous DOPC GUVs. The reason for that might be the lower density of the actin network that accommodates the stresses induced by actin polymerization without rupturing. Note that in all cases, the growth rate of the actin network is similar (*Fig-5D*) but slower than for DOPC GUVs (*Fig-5B*), consistent with the fact that the concentration of Ni-NTA lipids in our DOPC GUVs is 2.5 times higher.

Therefore, the distribution of lipid domains affects the structure of the actin network: a small number of large domains induces denser networks, discontinuous at the GUV scale, whereas a high number of small domains induces low-density continuous actin networks.

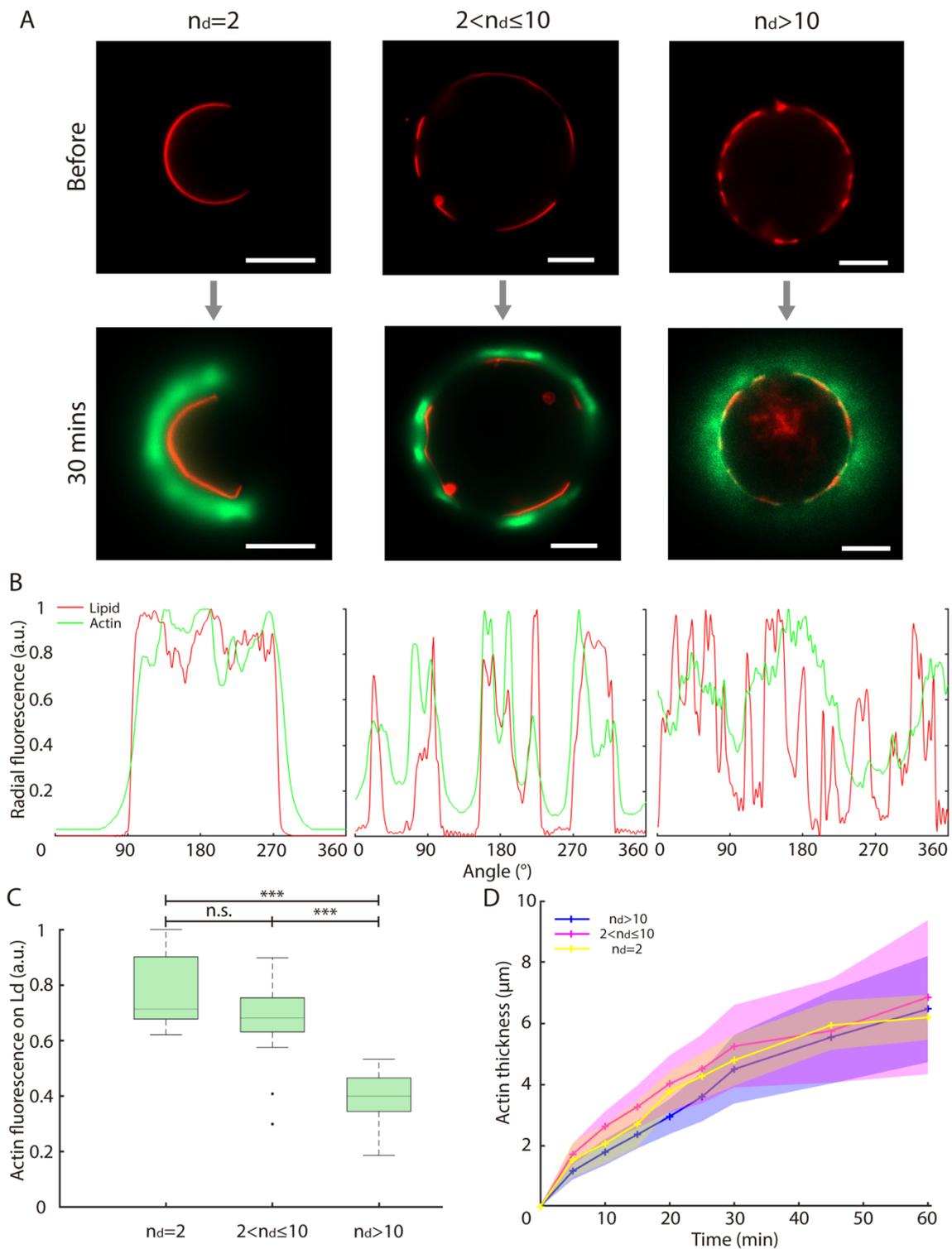

**Figure 5: Lipid domain number influences actin network structure**. (A) Actin polymerization on phase-separated GUVs with different domain numbers; Scale bars: 10 μm. (B) Normalized radial fluorescence of lipid and actin channels for the GUVs presented in *(A)* after 30 minutes of actin polymerization; (C) Normalized actin network fluorescence on Ld domains for the three categories presented in (A) ($n_d = 2$, N = 10 GUVs; $2 < n_d \leq 10$, N = 24 GUVs; $n_d > 10$, N = 21 GUVs); (D) Actin network thickness over time for the three categories (mean ± standard deviation, same GUVs numbers as in (C)). Statistical analysis was performed using Welch's t-tests.

## 4 Conclusions

In this work, we introduce a microfluidic platform based on micron-scale PDMS microtraps to study the dynamics of reconstituted actin networks interacting with GUVs. Our new experimental platform and protocol address key challenges in studying GUV remodeling. We characterize the loading of our microfluidic chambers with GUVs and actin solutions and show that we control the sequential introduction of actin and actin-binding proteins on immobilized GUVs. Then, we show that we can track populations of single and isolated GUVs over time from their initial state to generate statistically significant data.

We characterize the polymerization dynamics of branched Arp2/3 nucleated F-actin networks on homogeneous DOPC GUVs in the microfluidic chamber. We show that these experiments yield the same results as previously reported experiments on GUV coupled to actin networks, validating our experimental strategy.

Then, we use our approach to obtain novel results on phase-separated GUVs by first demonstrating how actin networks stabilize lipid domains by preventing their coalescence. Previous works have shown the lipid domain stabilization effect by dynamic DNA Line-Actants binding and FtsZ polymerization[43,59] but never by reconstituted actin networks.

Besides, we show that the actin network structure depends on the lipid domain number. GUVs with numerous lipid domains exhibit low-density actin networks covering the whole GUV, and GUVs with fewer lipid domains display denser networks strongly colocalized with the Ld domains. The number and density of domains are, therefore, a proxy to control the density and structure of the actin network. Moreover, these experiments thus show reciprocal crosstalk between lipid domains and actin networks: actin affects the dynamics of the domains, and domain number affects the network's structure.

When a lipid membrane presenting liquid-liquid phase-separation is brought below the critical mixing temperature, many small domains (below the optical resolution) nucleate. Then, the number of domains subsequently decays (or, equivalently, the average domain size increases) over time because of two mechanisms: domains coalesce when they come into contact by 2D diffusion within the membrane, and lipids migrate from small to large domains by Ostwald ripening (Figure 6A). Experimental observations have however demonstrated that Ostwald ripening is a slow process, and that domain coarsening is dominated by coalescence under conditions similar to those considered here[60,61]. The coalescence dynamics is therefore controlled by the domain diffusion coefficient, which can be estimated from the Saffmann-Delbruck equation: $D = \frac{k_B T}{4\pi \eta_m} \ln\left(\frac{\eta_m}{\eta_f r}\right)$ where $\eta_m$ is the membrane viscosity, $\eta_f$ the viscosity of the surrounding fluid, and r the domain size. In our experiments, as actin is polymerized on Ld domains, we consider the 2D diffusion of Ld domains (mostly DOPC) within a Lo (mostly SM and Chol) membrane. For Lo membranes, $\eta_m$ is on the order of 10$^{-8}$ Pa.m.s[62], and $\eta_f \simeq 10^{-3}$ Pa.s is the water viscosity. Therefore, a Ld domain of size 1 μm (the smallest domain size that we can resolve on our optical images) has a diffusion coefficient $D \cong 8 \times 10^{-14}$ m$^2$/s. We can estimate the time for this domain to be in contact with its nearest neighbor, $\tau \cong \frac{R^2}{4D}$ where $R^2$ is the typical surface on which the domain of interest diffuses. In our case, R is on the order of 5 μm (estimated from Fig.5), yielding $\tau \cong 80\ s$. This agrees with our observation that the domain number decreases over time scales of the order of minutes as shown in Fig. 4B and reported by others. Now, considering an Ld domain covered by an actin network, we propose that the actin network grows laterally as well as vertically over the lipid domain, thus forming an overhang that sterically hinders domain coalescence. Over time, the sizes of the actin networks become clearly larger than those of the Ld lipid domains from which they emerge, with an additional extension of around a micrometer. With the actin gel growth rate of 0.15 μm/min (or 2.5 nm/s) obtained from Fig.3C, during the typical minute-scale diffusion time $\tau$ required for two adjacent domains to come into contact, the gel grows by around 200 nm. This is more than the estimated mesh size of this type of actin networks, namely around 100 nm[63]. In 30 min, the time at which images of Fig. are captured, this corresponds to an actin layer of 4.5 μm. Therefore, each domain is likely surrounded by an actin corona that prevents its coalescence with others, explaining why we observe domain stabilization in the presence of actin. Note that this corona could also slow the domain's diffusion down.

In addition to its influence on coalescence, this effect also explains why GUVs with many small domains exhibit continuous actin networks covering the whole GUV. Smaller Lo domains (typically under 5 micrometers), do not nucleate actin filaments but display a higher actin fluorescent intensity (Supplementary Fig.5). This confirms that actin networks formed from adjacent Ld domains cover these small Lo domains, and come into contact to form an overall uniform actin network surrounding the whole GUV observed in Fig.5.

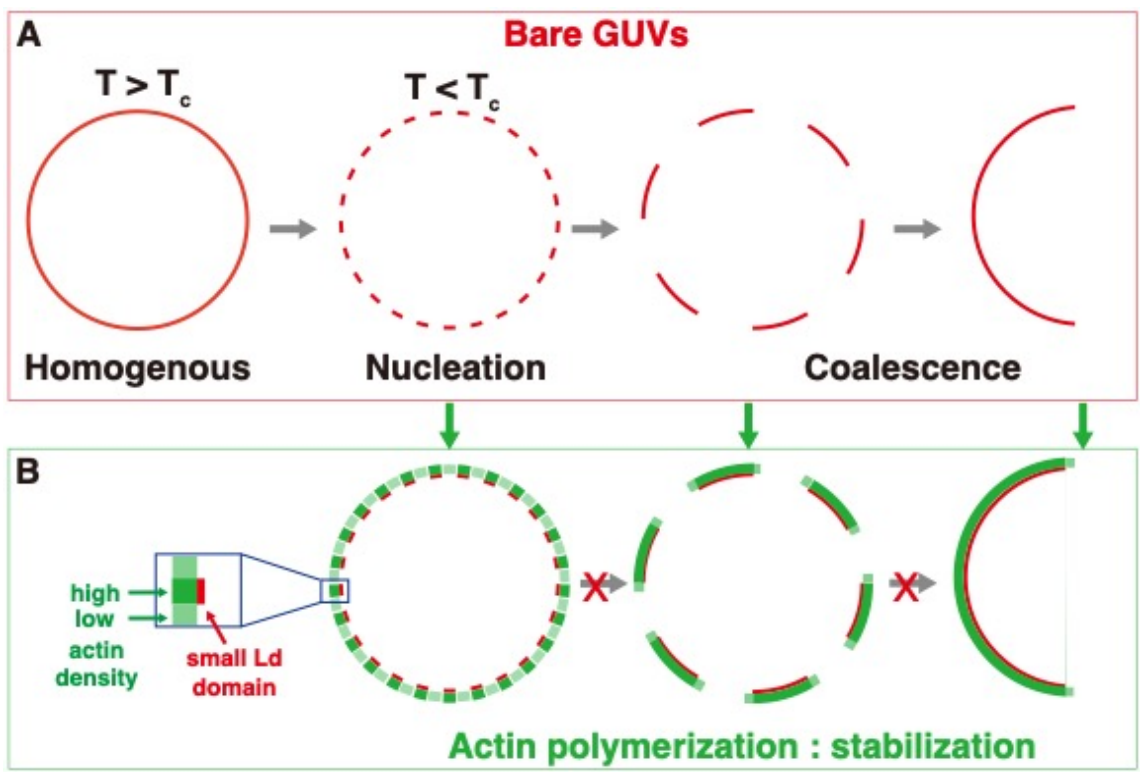

**Figure 6: Scheme of domain dynamics for bare GUVs and in the presence of actin networks.** (A) For bare GUVs, after domain nucleation, domains coarsen by coalescence. (B) Actin polymerization on phase-separated GUVs blocks domain coalescence by generating an actin corona extending around Ld domains. In the case of many small domains, a global continuous actin network surrounds the GUVs (bottom left case).

By enabling detailed investigation of actin-induced shape changes in GUVs, our approach allows further exploration using other NPFs, actin-binding proteins, or actin homologs to study the associated membrane remodeling and mimic the biochemical environment of living cells. Moreover, our microfluidic approach holds potential for broad applications beyond actin-GUVs assays, extending to any experiments investigating the remodeling of GUVs membrane by proteins, peptides, or other relevant compounds. As such, it paves the way for biophysical studies to deepen our understanding of membrane-associated biological functions.

**Author Contributions**

Conceptualization: Lixin Huang, Rogério Lopes-Dos-Santos, Martin Lenz and Clément Campillo. Methodology: Lixin Huang, Rogério Lopes-Dos-Santos, Jacques Fattaccioli and Clément Campillo. Experiments: Lixin Huang and Rogério Lopes-Dos-Santos. Fabrication: Lixin Huang, Rogério Lopes-Dos-Santos, Olek Maciejak and Jacques Fattaccioli. Sample: Lixin Huang and Rogério Lopes-Dos-Santos. Resources: Guillaume Lamour, Sid Labdi, Olek Maciejak, John Manzi, Jacques Fattaccioli, Michel Malo and Clément Campillo. Writing – original draft: Lixin Huang. Writing – review & editing: Lixin Huang and Clément Campillo.

**Conflicts of interest**

There are no conflicts to declare.

**Acknowledgments**

The authors are grateful for the financial support given by the China Scholarship Council (CSC, No. 202208070064) assigned to L.H and funding from the French Agence Nationale de la Recherche (ANR-18-CE13-0007-0 awarded to C.C.) and from Institut Universitaire de France (to C.C.). The authors thank Drs. Cécile Leduc, Antoine Jégou